\documentclass[5p,twocolumn]{elsarticle}

\usepackage{hyperref}
\usepackage{booktabs}
\usepackage{multirow}
\usepackage{graphicx}
\usepackage{subfigure}

\journal{Journal of \LaTeX\ Templates}

\begin{document}

\begin{frontmatter}

\title{Detecting Offensive Language on Social Networks: An End-to-end Detection Method based on Graph Attention Networks}


%\author{Zhenxiong Miao, Xingshu Chen, Haizhou Wang, Rui Tang, Zhou Yang, Wenyi Tang}
%\address{School of Cyber Science and Engineering, Sichuan University, Chengdu, 610207, China}
%\address{Cyber Science Research Institute, Sichuan University, Chengdu, 610207, China}


\author[1]{Zhenxiong Miao}

\author[1,2]{Xingshu Chen}

\author[1]{Haizhou Wang\corref{mycorrespondingauthor}}
\cortext[mycorrespondingauthor]{Corresponding author}
\ead{whzh.nc@scu.edu.cn}

\author[1]{Rui Tang}

\author[1]{Zhou Yang}

\author[1]{Wenyi Tang}

\address[1]{School of Cyber Science and Engineering, Sichuan University, Chengdu, 610207, China}
\address[2]{Cyber Science Research Institute, Sichuan University, Chengdu, 610207, China}

\begin{abstract}
The pervasiveness of offensive language on the social network has caused adverse effects on society, such as abusive behavior online. It is urgent to detect offensive language and curb its spread. Existing research shows that methods with community structure features effectively improve the performance of offensive language detection. However, the existing models deal with community structure independently, which seriously affects the effectiveness of detection models. In this paper, we propose an end-to-end method based on community structure and text features for offensive language detection (CT-OLD). Specifically, the community structure features are directly captured by the graph attention network layer, and the text embeddings are taken from the last hidden layer of BERT. Attention mechanisms and position encoding are used to fuse these features. Meanwhile, we add user opinion to the community structure for representing user features. The user opinion is represented by user historical behavior information, which outperforms that represented by text information. Besides the above point, the distribution of users and tweets is unbalanced in the popular datasets, which limits the generalization ability of the model. To address this issue, we construct and release a dataset with reasonable user distribution. Our method outperforms baselines with the F1 score of 89.94\%. The results show that the end-to-end model effectively learns the potential information of community structure and text, and user historical behavior information is more suitable for user opinion representation.
\end{abstract}


\end{frontmatter}

\section{Introduction}

In recent years, more and more offensive language appears in social networks, which leads to cyber violence and abusive behavior. The offensive language has become a serious concern for governments and social media platforms \cite{zampieri2019predicting}. There are several reasons why offensive language is proliferating in social networks. On one hand, users hide their identities in social networks easily. The anonymity of social networks increases the chances that users adopt aggressive behavior. On the other hand, social networks provide an environment where offensive language spread quickly \cite{fortuna2018survey}.

Usually, social platforms will adopt detection algorithms to find offensive language automatically and eliminate them. However, it is not easy to automatically identify these contents, especially the offensiveness is contained in the hidden rhetoric and irony \cite{mosca2021understanding}. Offensive language can be divided into explicit offensive language and implicit offensive language \cite{waseem2017understanding,KAPIL2020106458}. Explicit offensive language refers to language that is explicitly abusive in its meaning, such as racial discrimination. Implicit offensive language refers to language that does not directly imply abuse, and it is difficult to judge whether it is offensive without understanding the context \cite{waseem2017understanding,SHETH2021}.

For offensive language detection, existing research \cite{magu2017detecting,djuric2015hate,nobata2016abusive,davidson2017automated} have done some works based on text, confirming that n-gram features have good performance. Meanwhile, some researchers \cite{dadvar2013improving,waseem2016hateful,unsvaag2018effects} have combined user profile features such as age and gender with text features to improve the performance of detection. Recently, more and more methods based on deep learning were applied \cite{badjatiya2017deep,park2017one,yoon2018learning,LEHONG2021107504}. The BERT-based methods started to appear in offensive language-related tasks of International Workshop on Semantic Evaluation (SemEval)  \cite{liu2019nuli,dai2020kungfupanda,safaya2020kuisail}. In addition, some works \cite{mishra2018author,mishra2019abusive} have further studied the representation method of user features. They introduced community structure into offensive language detection, which significantly improved the detection performance.

However, there are limitations in existing methods using community structures. These methods \cite{mishra2018author, mishra2019abusive} deal with community structure independently, which seriously affects the effectiveness of user representations. This limits the performance of the offensive language detection model. On the other hand, the existing datasets are mainly collected by searching with the keyword in tweets, only considering the tweets containing explicit offensive words, such as \cite{waseem2016hateful}, \cite{davidson2017automated}, \cite{zampieri2019predicting}. Some abusive tweets that do not contain predefined keywords cannot be collected. This causes the existing models unable to learn relevant feature information and reduces the detection performance and generalization ability of the model in detecting implicit offensive language \cite{ribeiro2018characterizing}. Furthermore, some popular datasets may be affected by ``author bias" \cite{wiegand2019detection, arango2019hate}. This means that the classifier can be affected by the unbalanced distribution of users, which limits the generalization ability of the model. For the dataset of Waseem and Hovy \cite{waseem2016hateful}, more than 70\% of the sexist tweets originate from two users, and 99\% of the racist tweets originate from one user. Existing research shows that detection models based on such datasets could produce severe user overfitting \cite{arango2019hate}.

To solve these problems, we conduct the work in this paper with the following contributions.
\begin{itemize}	
	\item An end-to-end method based on community structure and text features are proposed for offensive language detection (CT-OLD). It uses Graph Attention Networks (GAT) \cite{velivckovic2017graph} to learn user representations from community structure and fine-tuned BERT \cite{devlin2019bert} to learn the text representations. The model learns these features at the same time. In addition, attention mechanisms and position coding are used for feature fusion. The CT-OLD achieves an F1 score of 89.94\% with our dataset. Moreover, our method has better robustness in different ratios of training and test set. The experimental results show that the end-to-end model significantly improves the current detection performance.
	\item A novel method is proposed to represent users with user historical behavior information. The user historical behavior information indicates the opinion tendency of users to post offensive language. This method reduces the number of parameters in the graph neural network and improves the performance of the model. For users whose historical behavior is unknown, four feature initialization methods are compared. Our method improves the F1 score by 2.57\% over the user representation method in \cite{mishra2019abusive}.
	\item A new dataset with community structure and reasonable user distribution is constructed \footnote{The dataset will be released on \par https://github.com/mzx4936/CT-OLD-Dataset}. A method that depend on the lexicon indirectly is used to construct the dataset. Specifically, we first collect the users and relationships rather than tweets. Then the tweets of the users are collected. We collect over 400,000 tweets and over 500,000 users. Then we annotate 12,780 tweets posted by 1260 users. The offensive tweets are posted by 220 users, accounting for 17.5\% of the users.
\end{itemize}

\section{Related Work}

Recently, research based on social networks has been increasing \cite{tang2019second, tang2021susceptible, tang2021learning, tang2020budgeted}. In this section, we summarize the research on offensive language detection in social networks in recent years. The methods in offensive language detection research are divided into two main categories: classical machine learning-based methods \cite{magu2017detecting, nobata2016abusive,dadvar2013improving,waseem2016hateful,unsvaag2018effects,razavi2010offensive,sood2012using,mishra2018author,salminen2018anatomy,grover2016node2vec}, and deep learning-based methods \cite{badjatiya2017deep, liu2019nuli, dai2020kungfupanda, safaya2020kuisail,mishra2019abusive, ribeiro2018characterizing, devlin2019bert, chen2020verbal, rajamanickam2020joint,  caselli2020hatebert,qian2018leveraging, kipf2016semi}.

\subsection{Classical machine learning-based methods}
Classical machine learning methods have been widely used in offensive language detection task. It is divided into two categories: methods based on text features \cite{magu2017detecting, nobata2016abusive,waseem2016hateful,razavi2010offensive, sood2012using,  salminen2018anatomy}, and methods combining text and user features \cite{dadvar2013improving, unsvaag2018effects, mishra2018author, grover2016node2vec}.
\subsubsection{Text-based feature methods}
Most of the text feature-based offensive language detection methods involve supervised text classification tasks, which generally requires feature engineering, such as text features, sentiment features, linguistic features. Text features are calculated by bag-of-words models or word embeddings. Sentiment features of the sentence are obtained by predefined word sentiments. Linguistic features refer to the statistical results of uppercase letters, lowercase letters, and punctuation marks. Using feature engineering methods, various classifiers can be supervised training.

Razavi et al. \cite{razavi2010offensive} constructed a lexicon containing insulting and abusive language. Assigning a weight to each word in the lexicon to indicate its impact, which effectively improved the accuracy of detection. Sood et al. \cite{sood2012using} pointed out that sometimes some words appearing in the blacklist would not cause offensiveness. At the same time, they realized that offensive language may hide offensive words through accidental or deliberate spelling errors, so edit distance was used to find potentially offensive words. Finally, a combination of Support Vector Machine (SVM), lexicon, and edit distance is used for classification. The integrated method achieved the highest F1 score compared to the single method. Moreover, since pronouns can be used to avoid detecting offensive terms, Magu et al. \cite{magu2017detecting} added known offensive word counterparts to their lexicon.

In addition, some studies used n-gram features for offensive language detection. Waseem and Hovy \cite{waseem2016hateful} compared word n-grams and character n-grams with Logistic Regression (LR) for classification. The results showed that character n-grams performed best. Nobata et al. \cite{nobata2016abusive} combined multiple features to detect offensive language. They combine features such as n-gram features obtained by text, linguistic features obtained by statistics, and distributed semantic features obtained by word2vec to detect offensive language. Through experiments on multiple datasets, they found that fusing multiple features greatly improves detection performance compared to single feature. Salminen et al. \cite{salminen2018anatomy} proposed an improved n-gram method. To improve the effectiveness of n-grams, they tried to weight n-grams using word frequency and TF-IDF. The classification was performed using LR, Decision Tree, Random Forest (RF), Adaboost, and SVM. Finally, the n-grams weighted using TF-IDF with SVM achieved 96\% of F1 score in binary classification.

\subsubsection{Methods of combining text features with user features}
Several researchers have combined personal characteristics such as age, gender, and activity history of the users in detecting offensiveness. Dadvar et al. \cite{dadvar2013improving} combined textual content and user-based features. They considered the history of user activity, the number of offensive appearances in previously posted content, and the age of the user. The addition of user features improved the F1 score by 4\%, with the highest contribution from the number of users' historical offensive language.
Unsvåg and Gambäck \cite{unsvaag2018effects} used user profile information (gender, geographic location, etc.), social network-based user information (number of followers and followees), and activity-based user information (number of favorites, etc.) as user features in combination with n-grams of text for classification. Their work showed that user features could effectively improve the performance of offensive language detection.

Offensive language often comes from users who share common stereotypes \cite{mishra2018author}, while existing offensive language detection methods completely ignore the structure of user communities. Mishra et al. \cite{mishra2018author} proposed a new method of using community information. They collected user and social relationship information on the basis of the Waseem and Hovy dataset \cite{waseem2016hateful}. The users were used as nodes and follower-followee relationships were used as edges to form an undirected social graph. Then the user node embeddings were obtained by node2vec \cite{grover2016node2vec} and text features were obtained by n-grams. Finally, the LR classifier was trained for classification, which greatly improved the performance of detection and obtained state of the art at that time. Their work illustrated the importance of using community-based information for offensive language detection.

Classical machine learning methods need to analyze the data and complex features manually, which is probably losing potential features. Therefore, more and more researchers have started to apply deep learning methods to offensive language detection.

\subsection{Deep learning-based methods}
Deep learning-based methods automatically represent text or users through deep learning models, greatly reducing the reliance on feature engineering. It is divided into the following two types of methods: the method of using text \cite{badjatiya2017deep, liu2019nuli, dai2020kungfupanda, safaya2020kuisail, devlin2019bert, chen2020verbal, rajamanickam2020joint,  caselli2020hatebert}, and the method of combining text information with user information \cite{mishra2019abusive, ribeiro2018characterizing,qian2018leveraging,  kipf2016semi}.
\subsubsection{Text-based embedding methods}
Recently, deep learning techniques have been applied to offensive language detection task. Such methods usually rely on word embeddings to obtain text representation and then train them by sequential or convolutional models. Badjatiya et al. \cite{badjatiya2017deep} proposed the methods based on deep learning models and compared them with machine learning methods. The authors compared two word vector initialization methods, random embedding and Glove embedding. After that, the classifiers of CNN, LSTM, and FastText were used to classification. The highest F1 score of deep learning models reached 83.9\%, while the highest F1 score of the baseline method using machine learning was 81.6\%. In addition, the method combining LSTM and GBDT achieves the highest F1 score of 93.0\%. In this method, the word vectors were trained by a deep neural network after random initialization of the word vectors, and then input into GBDT as features for classification.
Chen et al. \cite{chen2020verbal} considered that for short texts in social networks, it was difficult to fully learn the information of each word with a highly context-dependent method  such as word embeddings. Therefore, they proposed a method based on TF-IDF and CNN, which transformed the 1-dimensional vector obtained by TF-IDF into 2-dimensional matrix, allowing CNN to convolve the word features in a larger field. Compared with randomly initialized word embeddings, the accuracy rate was increased by 9\% and the AUC value was increased by 0.12.
Rajamanickam et al. \cite{rajamanickam2020joint} proposed a multi-task learning detection model for emotion and offensive language. They used offensive language detection as the primary task and emotion detection as the auxiliary task. The output of the auxiliary task was added to the primary task by encoding the text with BiLSTM. And the information flow was controlled by learnable parameters. When compared on two popular datasets, the method using multi-task learning improved F1 by 1\% over the single-task method.

With the rise of large-scale pre-trained language models, BERT \cite{devlin2019bert} has occupied an important position in natural language processing. In the SemEval-2019 Task 6 task, Liu et al. \cite{liu2019nuli} obtained first place in detecting whether a text is offensive (subtask A) by fine-tuning the BERT model for transfer learning only. Dai et al. \cite{dai2020kungfupanda} achieved better results by fine-tuning the BERT model for multi-task learning based on Liu et al. and experimenting on the same dataset.

Although the BERT model has achieved good performance on numerous NLP tasks, when it is applied to non-standard language types such as social network posts, the detection performance may fluctuate greatly due to its chaotic nature. Thus Caselli et al.\cite{caselli2020hatebert} proposed the HateBERT model, a retraining BERT model for detecting offensive language in English. The model was trained on a large-scale Reddit English comment dataset that included posts from communities banned for being offensive, abusive, and promoting hate speech, containing a total of 1,492,740 messages. The authors compared with the generic BERT model on three English general-purpose datasets, and the F1 score of HateBERT were all higher than those of generic BERT.
Safaya et al. \cite{safaya2020kuisail} combined BERT and CNN to obtain better detection results than BERT. The authors split the tweets obtained from Twitter, transforming Hashtags into raw text, e.g., \emph{``\#SomeHashtagText"} into \emph{``Some Hashtag Text"}. Subsequently, the text was split by Wordpiece, a BERT splitter, and the text was fed into a twelve-layer BERT model. The output context embeddings of the last four hidden layers of BERT were provided to the CNN to maximize the semantic knowledge in BERT. Finally, the output of the CNN was used for classification. Experiments were performed on three language datasets (Arabic, Greek, Turkish) and the F1 score was improved by 1\% over BERT alone.

\subsubsection{Methods of combining text information with user information}
The user information provides information that is difficult to obtain at the textual granularity. Several researchers have attempted to combine user information with textual information.

Qian et al. \cite{qian2018leveraging} proposed a method for offensive language detection using intra-user and inter-user representations. They collected and analyzed users' historical tweets to model intra-user tweets. For inter-user tweets, they could be collected from the dataset by locally sensitive hashing (LSH). Adding the inter-user representation and intra-user representation improved the F1 score by 7.1\% over the baseline using BiLSTM.
Mishra et al. \cite{mishra2019abusive} used GCN \cite{kipf2016semi} to learn user embeddings and combined with the bag-of-words representation of the text for offensive language detection. They constructed a heterogeneous graph of users and tweets. The user representation was learned through two layers of GCNs, with the labels of the tweet nodes as the targets of GCN learning. The output of the first layer of GCNs was used as the user embedding. After that, LR was applied to detection with the user embeddings and the bag-of-words representation of the text. This heterogeneous graph representation was able to model the community structure and the linguistic behavior of authors in the community. Therefore, their method achieved an F1 score of 85.4\%, refreshing the state of the art of offensive language detection. 

Due to the noise of social network texts and the subjectivity of offensive language, the dataset faced shortcomings in construction. There might be an oversimplification in offensive language detection. Therefore, Ribeiro et al. \cite{ribeiro2018characterizing} turned to identifying whether users were offensive or not.  They collected over 100,000 users and labeled 4,972 of them. For each user, the authors used the number of followers and followees, the number of favorites, centrality measurements (including betweenness, eigenvector centrality, and the in/out degree of each node), and the tweet features obtained by GloVe. The results of the three methods of GradBoost, AdaBoost, and GraphSage for classification were compared.  The semi-supervised node embedding algorithm GraphSage was found to outperform the other methods, with a 15\% improvement in F1 in the task of detecting offensive users.

In existing studies, n-grams are effective for offensive language detection task. Therefore, classical machine learning methods mostly combine n-grams with other features. With the development of deep learning, especially the emergence of BERT, researchers apply word embeddings to text representations rather than complex feature engineering. During the research on offensive language detection task, the question of how to use user features has been explored. Some researchers applied personal information such as gender and age to improve detection performance. Furthermore, the relationships between users have been taken into consideration. The structure of communities was composed into graphs for graph embedding learning. 
In the past, user features were obtained after the user model was trained. Then user features are combined with text features. However, the user features obtained by this method may not work well for the offensive language detection task. Meanwhile, the ``author bias" in the existing datasets also limits the generalization ability of the model. In our work, we construct the dataset with a method that relies on lexicon indirectly. The user historical behavior feature \cite{dadvar2013improving} is used to represent the opinion of users. Finally, the end-to-end model is used to learn user features and text features for classification.

\section{Methodology}

\begin{figure*}[htbp]
	\centering
	\includegraphics[scale=0.45]{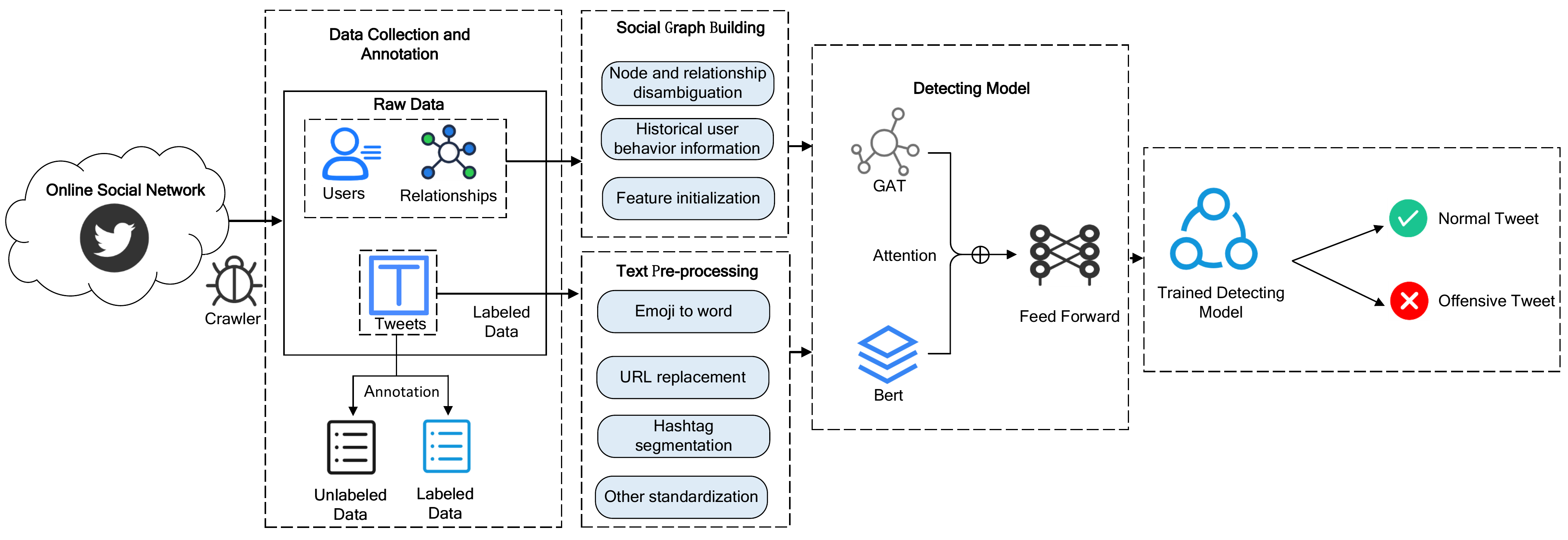}
	\caption{The architecture of the end-to-end offensive language detection methods}
	\label{architecture}
\end{figure*}

In this section, we will introduce the data collection and annotation module, text pre-processing module, social graph construction module, and detection model module. The framework of this paper is shown in Figure \ref{architecture}.

\begin{enumerate}[(1)]
	\item Data collection and annotation module: This module is responsible for collecting users, relationships, and tweets from Twitter to provide data support for other modules. In this module, a web crawler is developed to collect data from Twitter. Then the data is annotated according to the proposed seven annotation criteria.
	\item Social graph construction module: This module processes the collected user and social relationship into a social graph. It takes users as nodes, social relationships as edges, and historical behavioral information as node features to construct a directed isomorphic social graph.
	\item Text pre-processing module: This module is used to pre-process the collected tweets. In this module, the emojis in tweets are replaced by words with the same meaning. The hashtag is reduced to words or phrases. In addition, URLs, usernames, etc. are identified uniformly.
	\item Detection model: We detect offensive language by an end-to-end model that combines the text representation with community structure. The model is based on the GAT, the BERT, and the attention mechanism.
\end{enumerate}

\subsection{Dataset}
\subsubsection{Collection and annotation}
\paragraph{Data discuss}
The widely used datasets, including Waseem and Hovy \cite{waseem2016hateful}, Davidson et al. \cite{davidson2017automated}, and Zampieri et al. \cite{zampieri2019predicting}, start collecting through the tweets containing discrimination-related words such as religion, gender, and minority. Such a way of collecting the data would make the dataset have a much higher percentage of offensive language. However, it is difficult to collect tweets with more subtle offensive features. In addition, the datasets such as Waseem and Hovy \cite{waseem2016hateful} have severe ``author bias". More than 70\% of the sexist tweets originate from two users, and 99\% of the racist tweets originate from one user \cite{wiegand2019detection}. This bias affects the generalization ability of the model. In this case, Ribeiro et al. \cite{ribeiro2018characterizing} suggest collecting users firstly, and then collecting their tweets. Table \ref{dataset} shows the percentage of offensive language in English language datasets.

\renewcommand{\arraystretch}{1.5}
\begin{table}[htbp]
	\centering
	\fontsize{7}{8}\selectfont
	\caption{The statistics of current english datasets}	
	\addtolength{\leftskip} {-0.5cm} 
	\setlength{\tabcolsep}{0.2mm}{
	\begin{tabular}{@{}ccccc@{}}
		\hline
		\textbf{Dataset} &
		 \textbf{\begin{tabular}[c]{@{}c@{}}The number\\ of offensive \\ tweets\end{tabular}} &
		\textbf{\begin{tabular}[c]{@{}c@{}}The number\\ of tweets\end{tabular}} &
		\textbf{\begin{tabular}[c]{@{}c@{}}The ratio\\ of offensive \\ tweets\end{tabular}} &
		\textbf{Data type} \\ \hline
		Waseem and Hovy\cite{waseem2016hateful}      & 5,355 & 16,914 & 31.66\% & tweets  \\
		Davidson et al.\cite{davidson2017automated}  & -     & 24,802 & 76.00\% & tweets  \\
		Zampieri et al.\cite{zampieri2019predicting} & 4,640 & 14,100 & 32.91\% & tweets  \\
		Mandl et al.\cite{mandl2019overview}         & 4,456 & 7,005  & 63.61\% & tweets  \\
		Founta et al.\cite{founta2018large}          & -     & 80,000 & 17.50\% & tweets  \\
		Mishra et al.\cite{mishra2018author}         & 5,355 & 16,914 & 31.66\% & \begin{tabular}[c]{@{}c@{}}tweets, users,\\ relationships\end{tabular} \\
		Ours                                                          & 1,009 & 12,780 & 7.90\%  & \begin{tabular}[c]{@{}c@{}}tweets, users,\\ relationships\end{tabular} \\ \hline
	\end{tabular}}
	\label{dataset}
\end{table}

\paragraph{Data collection}
 Based on the above discussions, we want to build a dataset that the distribution of users and tweets is closer to the real-world environment. In most existing datasets, tweets are collected by keyword \cite{ribeiro2018characterizing}. Specifically, they first collect tweets using keywords from the lexicon. Then they collect the authors of the tweets and the relationships of the authors. However, they can't collect the tweets without keywords from the lexicon. In our method, we first collect the users based on the lexicon. Then we collect their relationships and tweets. This method will not lose any tweets.

 \begin{table}[htbp]
 	\centering
 	\fontsize{7}{8}\selectfont
 	\caption{Raw data and labeled data}
 	\begin{tabular}{@{}ccccc@{}}		
 		\hline
 		\textbf{Data Type}    & \textbf{Raw Data} & \textbf{Labeled Data} \\ 
 		\hline
 		Tweets       & 443,400   & 12,780        \\
 		Users        & 387,680   & 1,260         \\
 		Relationship & 559,735   & 8,877         \\ \hline
 	\end{tabular} 	
 \end{table}

Finally, our program collected 443,400 tweets, 387,680 users, and 559,735 relationships from Twitter. These tweets were posted from January 2018 to March 2021. Details of data collection are as follows.
\begin{enumerate}[(1)]	
	\item Instead of using offensive words, we construct lexicon by topic-related words. Then the users and the social relationships are collected with this lexicon. Specifically, the first-degree and second-degree friends and tweets are collected. Due to a large number of second-degree friends, we only collect tweets from first-degree friends. Finally, raw data includes 443,400 tweets, 387,680 users, and 559,735 relationships.
	\item In the raw data, we randomly select users and expand them to second-degree nodes. Then the users without tweets and social relationships are removed. Meanwhile, we focus on English offensive language detection and filter the others. After the tweets are annotated, only the users with annotated tweets and their relationships are retained. Finally, 1,260 users, 8,877 relationships, and 12,780 tweets are obtained. Among them, the number of offensive tweets is 1,009, accounting for 7.90\%.	
\end{enumerate}

\paragraph{Annotation}
The offensiveness of tweets includes racism, sexism, geography, etc. \cite{fortuna2018survey}. The offensiveness of many tweets is not obvious, so the determination of offensive tweets is a difficult task. Therefore, we need a clear discriminant to guide us in labeling tweets. Our dataset is annotated using the following criteria. The criteria refer to the annotation recommendations of Waseem et al.\cite{waseem2016hateful} and Zampieri et al.\cite{zampieri2019predicting}. Tweets that meet one of the following are labeled as offensive.

\begin{enumerate}[(1)]
	\item Criticizing or satirizing the target without well-founded arguments \cite{waseem2016hateful}.
		
	\item Abusing directed at individuals and groups \cite{zampieri2019predicting}.
		
	\item Blatant misrepresentation of facts or attempting to misrepresent the view of the subject with unfounded claims \cite{waseem2016hateful}.
		
	\item Negative stereotyping of an object.
		
	\item Discrimination against an aspect of the subject.
		
	\item Criticism of negative facts with a strong subjective negative bias.
		
	\item Causing discomfort.
\end{enumerate}
	
We label the tweets offensively and non-offensively, and this labeling is done by three graduate students. Two graduate students label the same tweet, and if the results are different, the third graduate student will make a judgment.

\subsubsection{Dataset analysis}
To analyze the user distribution and community distribution of the dataset, we perform the following statistics.

\begin{figure}[htbp]
	\centering
	\includegraphics[scale=0.6]{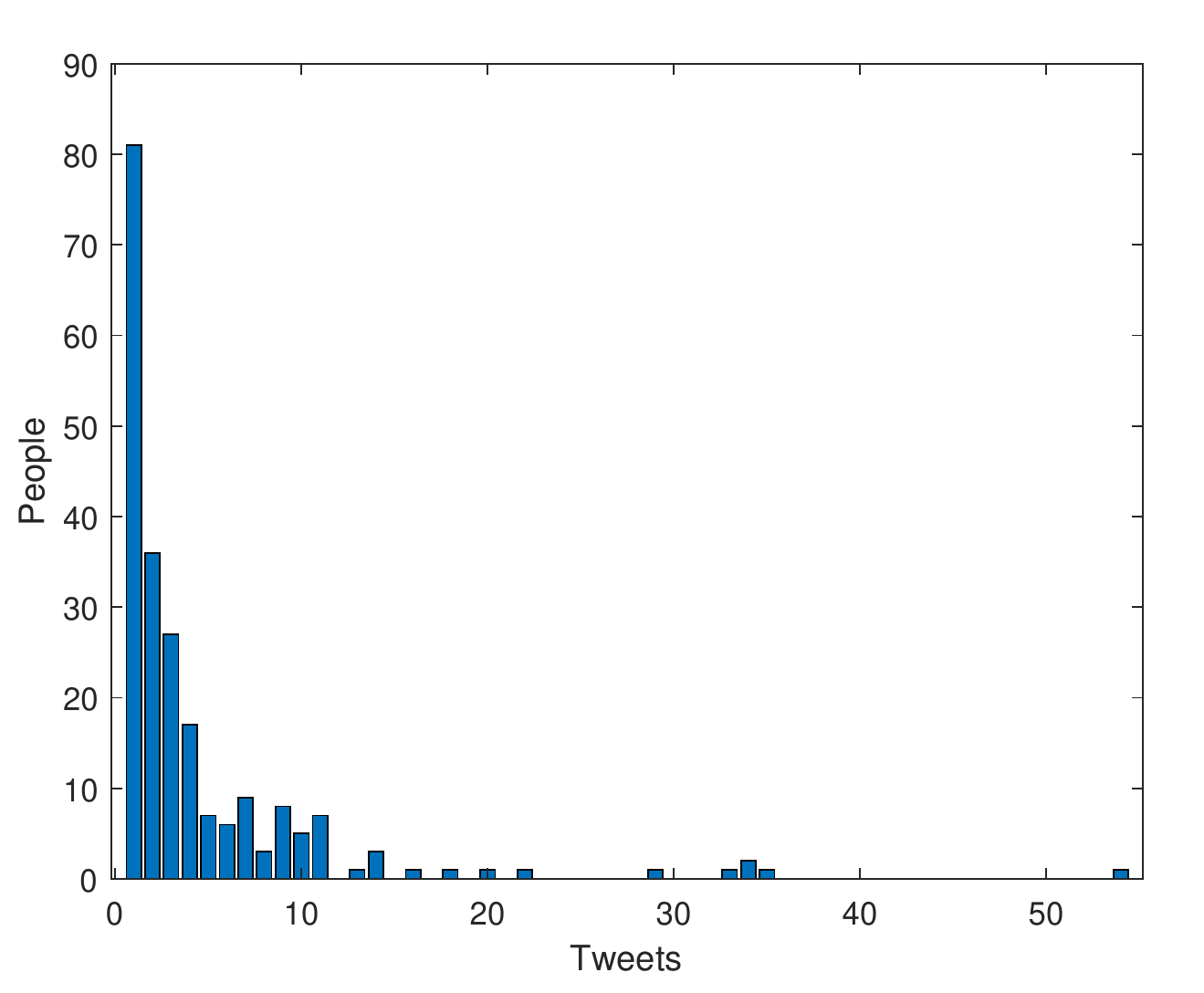}
	\caption{Distribution of users of offensive language}
	\label{bar}
\end{figure}

\paragraph{User analysis}
Among the 1,260 users, the median number of labeled tweets for each user is 11. There are 1,009 offensive tweets in the dataset, and these tweets are posted by 220 users. The distribution is shown in Figure \ref{bar}.

The horizontal axis of the figure above shows the number of offensive tweets posted by each user. The vertical axis shows the number of users who posted offensive tweets. As we can see, the majority of users posted between 1 and 11 tweets. The largest number of offensive tweets posted by one user is 54, accounting for 5.40\% of the total. The data distribution is more reasonable than the existing datasets  \cite{mishra2018author, wiegand2019detection}. The offensive tweets are posted by a small subset of users on social networks.

\paragraph{Community analysis}
All users are divided into 22 communities by the algorithms of Blondel et al. \cite{blondel2008fast} and Lambiotte et al.  \cite{lambiotte2008laplacian}. We show the user distribution of communities with the number of nodes greater than one. The communities are sorted by the proportion of offensive nodes, as shown in Table \ref{community}.

\renewcommand{\arraystretch}{1.5}
\begin{table}[htbp]
	\centering
	\fontsize{7}{8}\selectfont
	\caption{Statistics on community distribution of offensive users}
	\setlength{\tabcolsep}{0.5mm}{
	\begin{tabular}{@{}ccccc@{}}
		\hline
		\multirow{3}{*}{\textbf{Community id}} &
		\multirow{3}{*}{\textbf{\begin{tabular}[c]{@{}c@{}}The number of \\ nodes \\ in community\end{tabular}}} &
		\multirow{3}{*}{\textbf{\begin{tabular}[c]{@{}c@{}}The number of \\ offensive nodes \\ in community\end{tabular}}} &
		\multirow{3}{*}{\textbf{\begin{tabular}[c]{@{}c@{}}The percentage of \\ offensive nodes \\ in community\end{tabular}}} \\
		&     &    &         \\ 
		&     &    &         \\ \hline
		1 & 250 & 70 & 27.97\% \\
		2 & 241 & 15 & 6.22\%  \\
		3 & 219 & 55 & 25.10\% \\
		4 & 201 & 34 & 16.90\% \\
		5 & 130 & 15 & 11.52\% \\
		6 & 88  & 6  & 6.81\%  \\
		7 & 75  & 20 & 26.63\% \\
		8 & 23  & 1  & 4.34\%  \\
		9 & 21  & 1  & 4.75\%  \\ \hline
	\end{tabular}}
	\label{community}
\end{table}

\begin{figure*}[htbp]
	\centering	
	\subfigure[The users in communities 1, 3, 7]{
		\includegraphics[width=5.5cm]{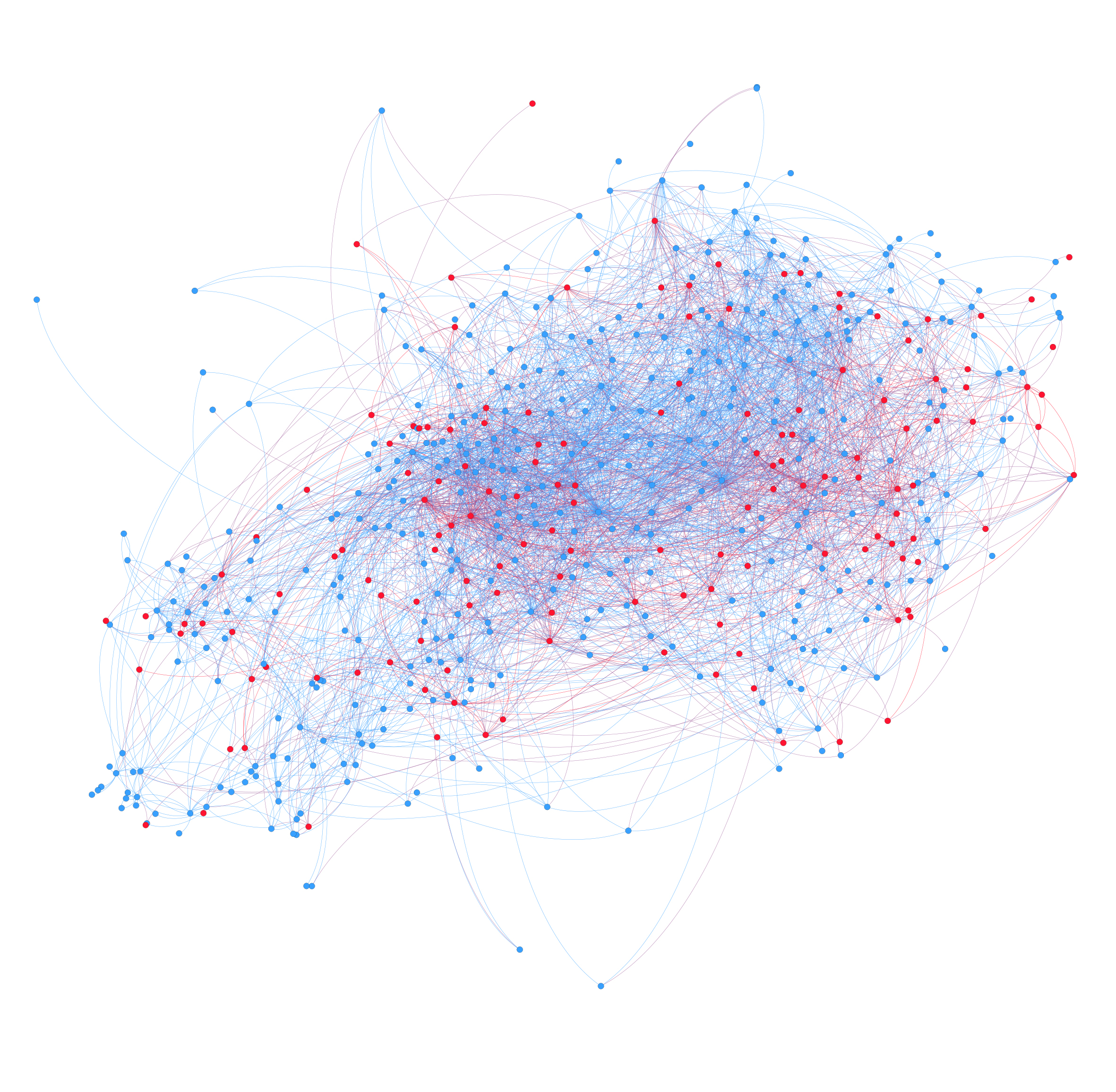}	
		\label{off}	
	}
	\quad
	\subfigure[The users in other communities]{
		\includegraphics[width=5.5cm]{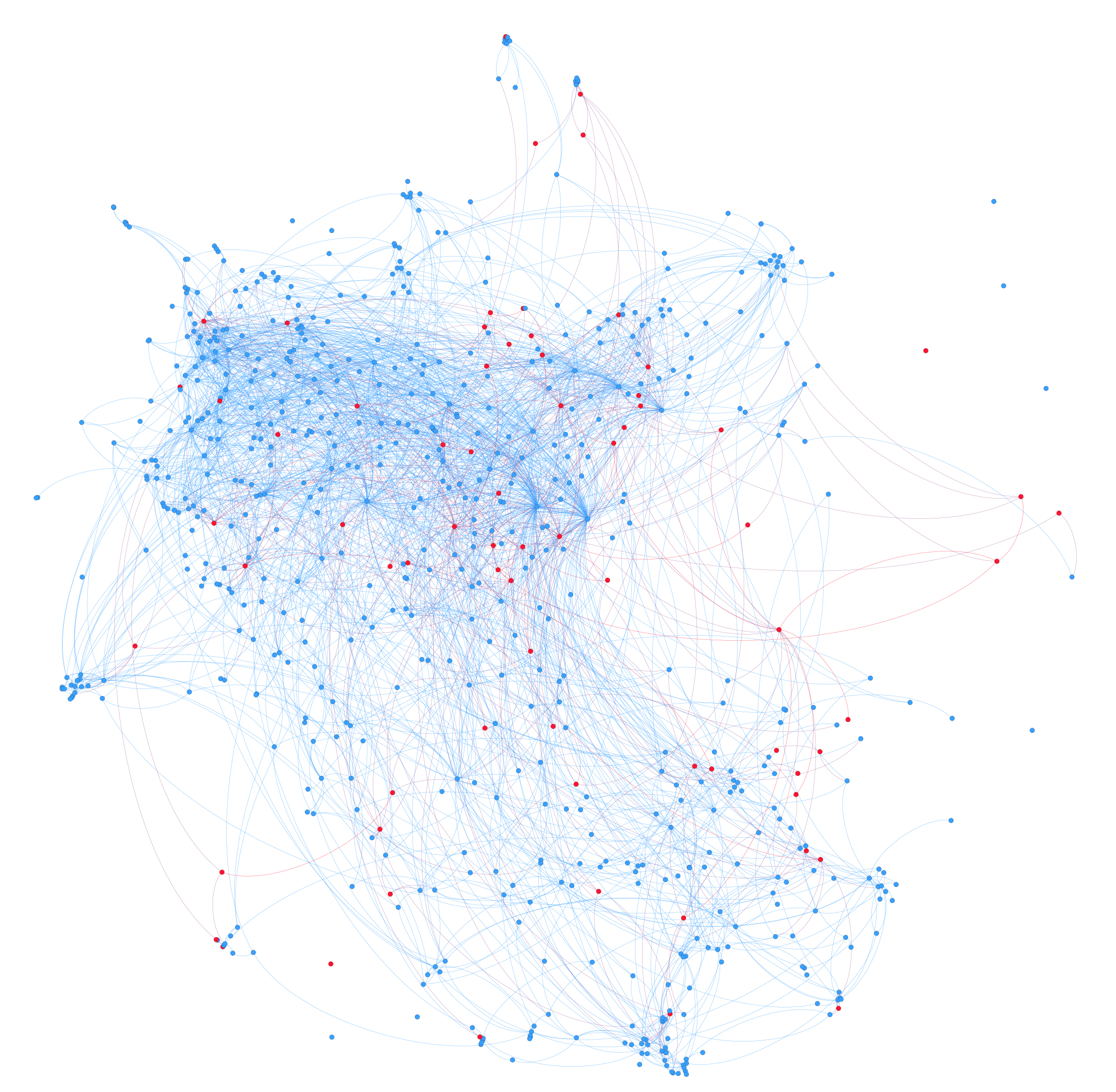}
		\label{notoff}
	}
	\quad
	\subfigure[All users]{
		\includegraphics[width=5.5cm]{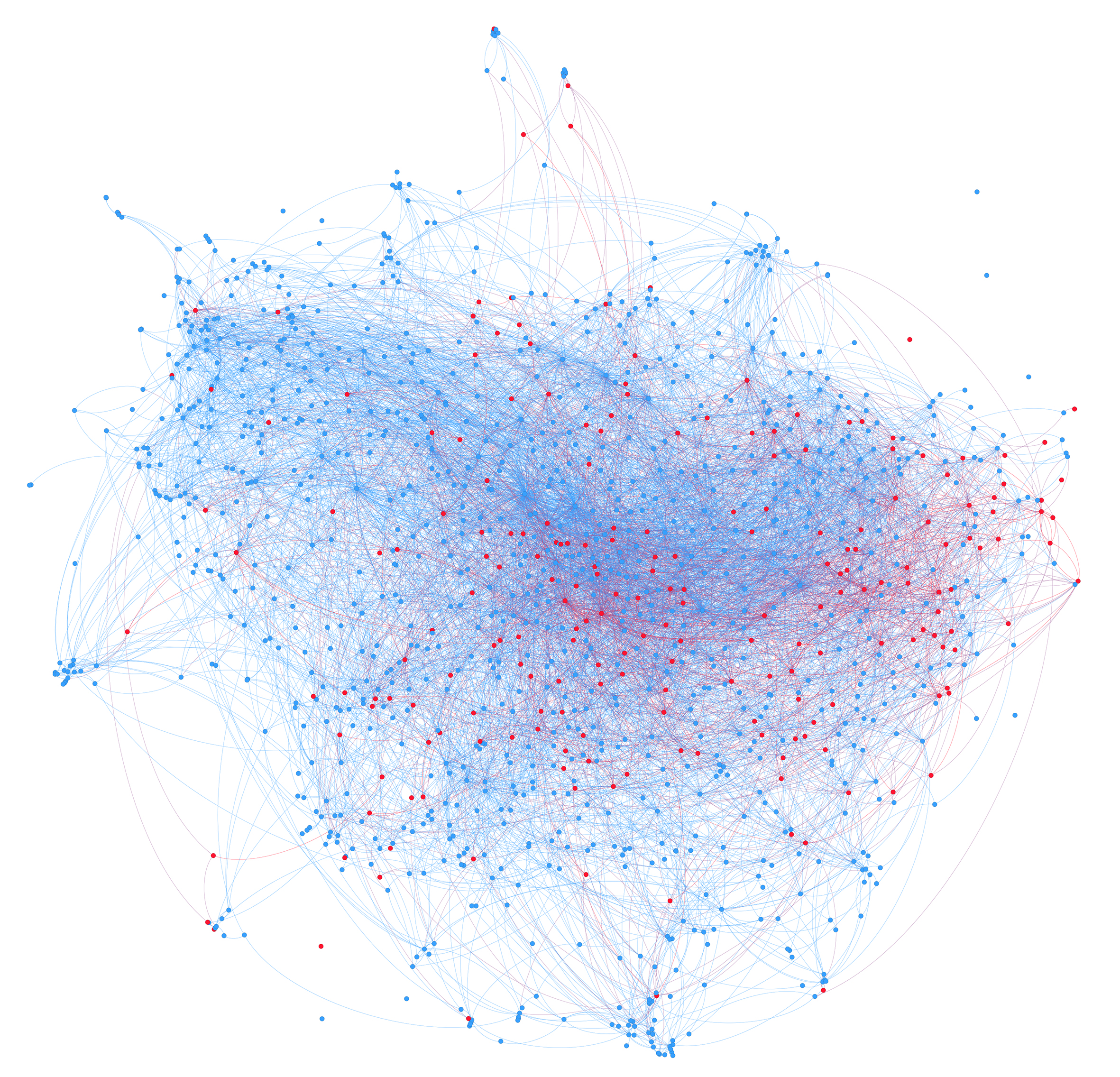}
	}
	\caption{The distribution of users in the communities}
	\label{social graph}
\end{figure*}

The Percentage of offensive nodes shows the ratio of offensive nodes in the total nodes of each community. More than 20\% of users in communities 1, 3, and 7 have posted offensive language, and they are shown in Figure \ref{off}. Figure \ref{notoff} shows the users of other communities. The blue nodes are users who have not posted offensive language. And the red nodes are users who have posted offensive language.

Combined with the discussions in dataset analysis, we draw the following conclusions.
\begin{enumerate}[(1)]
	\item Users who have posted offensive language are likely to continue to post offensive language.
	\item Users who post offensive language congregate in some communities. These communities contain the majority of users who post offensive language.
\end{enumerate}

\subsection{Social graph building}
The online social network is characterized by a complex network in which users are represented as nodes and the follower-followee information as edges \cite{TANG2020105598,TANG2022108651,9603289}. We construct a directed social graph. The direction of the edge is from the follower to the followee. If user $x$ follows user $y$ then an edge will be generated that points from $x$ to $y$. The attributes of the graph nodes only include information about the user historical behavior.

The offensiveness of users is associated with user historical behavior information rather than community structure \cite{mishra2019abusive}. Therefore, we use user historical behavior information to represent the offensiveness of users. The ratio of non-offensive language and offensive language in the training data is counted as user historical behavior information. For example, the user posted three offensive tweets and five non-offensive tweets, and the user historical behavior information is $(3, 5)$. It is used to indicate the opinion of this user to post offensive language. This feature is a node attribute on the social graph. If a user has no tweets, the default value $(1,1e-6)$ is used. Training model with this social graph allows the user features to represent the offensiveness of that user.

\begin{figure}[htbp]
	\centering
	\includegraphics[width=7cm]{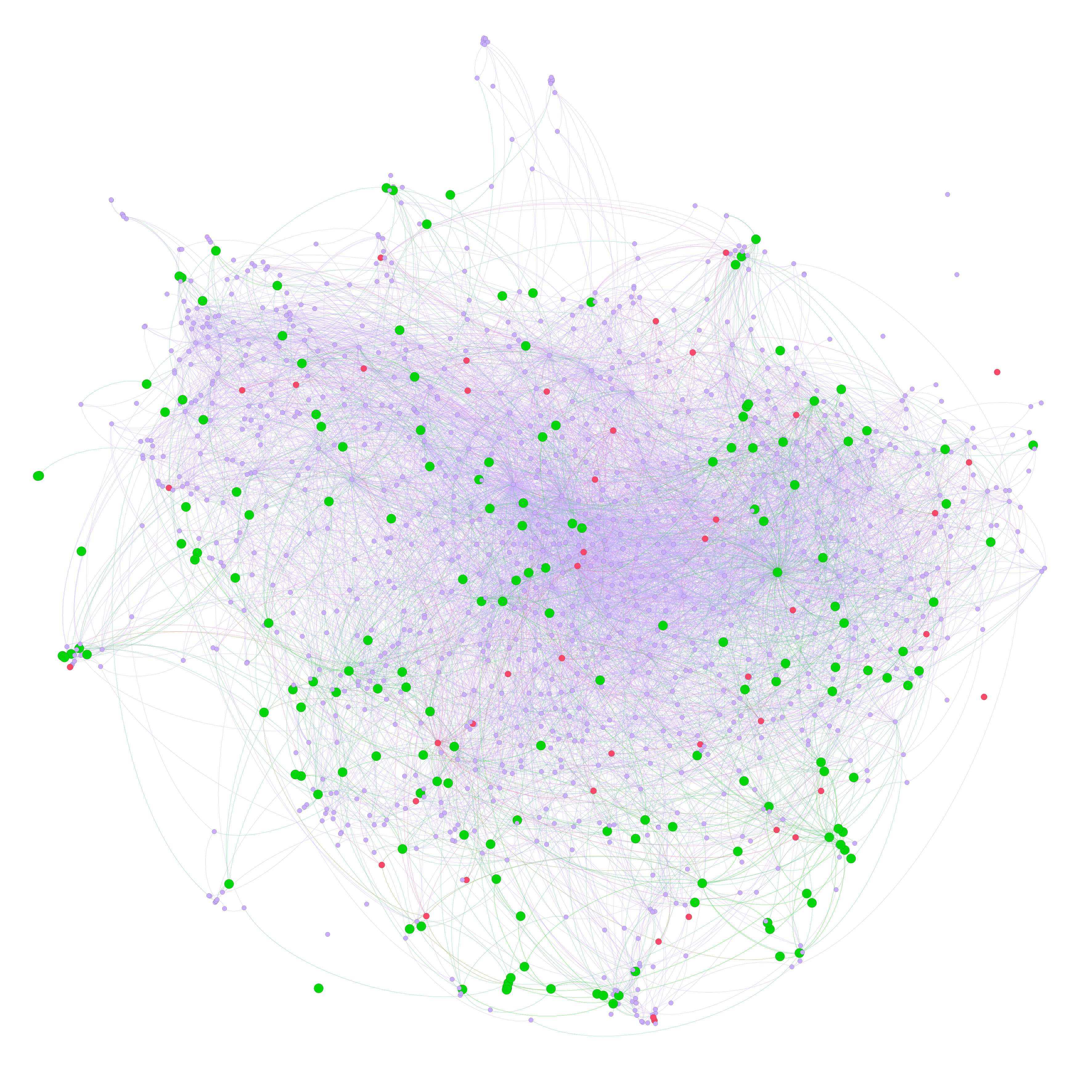}
	\caption{Training and test graph}
	\label{train-test}
\end{figure}

In this paper, the built graph consists of 1,260 nodes with 8,878 edges. We add self-loops to each node. The nodes have an average degree of 6.66. During the model prediction, the labels of tweets and user historical behavior information of the users are unknown. Therefore, the relevant information of the test set is removed from the social graph. The social graph we used for training and testing is shown in Figure \ref{train-test}. The red node indicates that the user's tweets are in the test set. The green node indicates that the user's tweets are in the training set. The purple node indicates that both the training set and the test set contain its tweets.

\subsection{Text pre-processing}
The processing of this part is directly related to the effectiveness of the offensive language detection task. We process the text from the following aspects.
\begin{enumerate}[(1)]
\item \textbf{Emojis.}
	Tweets contain a large number of emojis. Emojis play an important role in the sentiment tendency of tweets. If the emojis are removed directly, this part of sentiment information will be lost. We use Python's third-party libraries emoji\footnote{https://github.com/carpedm20/emoji} and ekphrasis\footnote{https://github.com/cbaziotis/ekphrasis} \cite{baziotis-pelekis-doulkeridis:2017:SemEval2} to convert emojis into words with similar meanings.
\item \textbf{URLs.}
	To avoid the impact of URLs on semantics of tweet embeddings, we replace all URLs with \emph{``http"}  \cite{dai2020kungfupanda}.  For example, \emph{``Refinancing during Covid-19. https://t.co/EzjVxwoLq7"} into \emph{``Refinancing during Covid-19. http"}
\item \textbf{Hashtag segmentation.}
	To utilize the textual information in hashtags, they are converted into phrases. This is done using the Python third-party library ekphrasis, e.g. \emph{``\#SomeHashtagText"} into \emph{``Some Hashtag Text"}.
\item \textbf{Other standardization.}
	Tweets containing user references, email addresses, time, currency, date, etc. may have an impact on the semantics of tweet embeddings. Thus, we convert these uniformly to \emph{``$<$category$>$"}, e.g. \emph{``@Kevin"} to \emph{``$<$user$>$"}, which uses ekphrasis.
\end{enumerate}

\subsection{Then end-to-end model}
\begin{figure*}[htbp]
	\centering
	\includegraphics[scale=0.55]{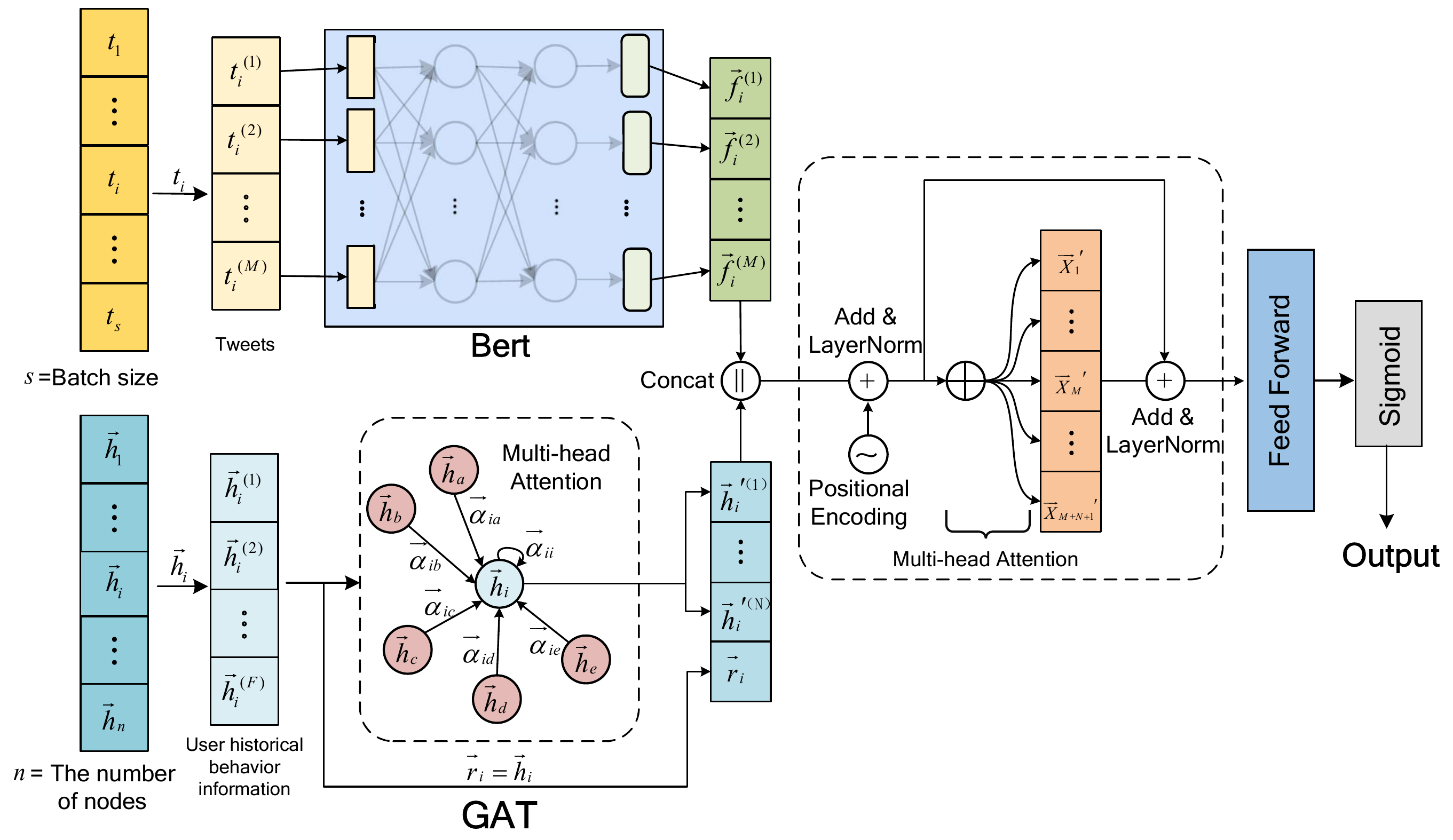}
	\caption{Structure of the end-to-end offensive language detection model}
	\label{model}
\end{figure*}

We design the end-to-end model for offensive language detection. The model consists of graph attention network layer, BERT, attention layer, and feed-forward layer. The learning rate of the GAT layer is set to 1e-2 and the rest are set to 5e-5. The parameters are initialized by the Xavier normal distribution \cite{glorot2010understanding}. In addition, the Adam optimizer \cite{kingma2015adam} is used as optimization. The structure of this model is shown as Figure \ref{model}.

\subsubsection{GAT layer}
Graph Attention Networks were proposed by Veličković et al. \cite{velivckovic2017graph}  in 2018. GAT uses the attention mechanism to assign different weights to different nodes, which is suitable for graph representation on social networks. The social graph is represented as $G=\left(V, E\right)$, $V$ is the set of user nodes and $E$ is the set of edges. The node features are represented as $h=\{\vec{h}_{1},\vec{h}_{2},...,\vec{h}_{n}\},\vec{h}_{i} \in R^{F}$, where $n$ is the number of nodes and $F$ is the dimensionality of the node features. We can obtain a high-level representation of the feature $h$ by the following linear transformation.

\begin{equation}
	z_{i}^{(l)}=W^{(l)}h_{i}^{(l)}	
\end{equation}

Where $W$ is the weight matrix, $h_{i}$ is the feature of each node, and $z_{i}$ is the transformed feature expression. The $(l)$ represents that this formula is the matrix calculation of layer $l$. 

For node $i$, the correlation coefficient is calculated by node $i$ and its one-degree neighbor $j$. The nonlinear function LeakyReLU is used as the activation function to obtain the attention scores. After that, the attention scores are normalized by softmax.

\begin{equation}
	e_{ij}^{(l)}=LeakyReLU(\vec{a}^{(l)^{T}}(z_{i}^{(l)}|z_{j}^{(l)}))	
\end{equation}

\begin{equation}
	a_{ij}^{(l)}=\frac{\exp (e_{ij}^{(l)})}{\sum_{k \in N_{(i)}}\exp (e_{ik}^{(l)}) }	
\end{equation}

Where $a$ is a single-layer feedforward neural network and $e_{ij}$ is the attention scores. The $a_{ij}$ is the normalized attention coefficient. 

After that, the attention coefficient is used to perform linear transformation with the corresponding node features. Finally, it is passed through the activation function as the output.

\begin{equation}
	h_{i}^{(l+1)}=\sigma(\sum_{j \in N_{(i)}} \alpha_{ij}^{(l)}z_{j}^{(l)})
\end{equation}

User representations are obtained by the multi-head attention mechanism, which stacks the attention of each head. In addition, to reduce the impact of the network on the original features, we perform a linear transformation on the original features and then stack them with the features calculated by attention. The formula is as follows.

\begin{equation}
	h_{i}^{(l+1)}=(||_{k=1}^{K}\sigma(\sum_{j \in N_{(i)}} \alpha_{ij}^{k} W^{k} h_{j}^{(l)}))||(Wh_{j}^{(l)})
\end{equation}

This part uses a single-layer graph attention network with eight attention heads and the hidden layer size is set to 768. The output of this layer is represented as  $H=\{H_{1}, H_{2}, \ldots, H_{n}\}$, where $n$ is the number of users. For the user $i$, $H_{i}=\left\{\vec{h}_{i}^{\prime(1)}, \vec{h}_{i}^{\prime(2)}, \ldots, \vec{h}_{i}^{\prime(N)}, \vec{r}_{i}\right\}$, and $N$ is the number of attention heads. The $\vec{r}_{i}$ is the residual, $\vec{r}_{i}=\vec{h}_{i}$ .

\subsubsection{BERT}
Google released Bidirectional Encoder Representation from Transformer (BERT) \cite{devlin2019bert} based on the Encoder of Transformer \cite{vaswani2017attention} in 2018 with improved performance for several NLP tasks. The BERT uses a large-scale corpus for pre-training. The model can be fine-tuned for downstream tasks when performing specific tasks.

This part mainly consists of the BERT Base model. The tweets are fed into the pre-trained BERT model. Then they are passed through 12 layers of self-attention to obtain the embeddings of the context. The attention dropout is set to 0.5 and the hidden dropout is set to 0.1. Meanwhile, texts longer than 512 are truncated. The input texts are denoted as $T=\left\{t_{1}, t_{2}, \ldots, t_{s}\right\}$, and $s$ is the number of texts in the current batch. For a text, it is expressed as $t_{i}=\left\{t_{i}^{(1)}, t_{i}^{(2)}, \ldots, t_{i}^{(M)}\right\}$, and $M$ is the length of the text. The output vectors are denoted as $F=\left\{F_{1}, F_{2}, \ldots, F_{n}\right\}$, where  $F_{i}=\left\{\vec{f}_{i}^{(1)}, \vec{f}_{i}^{(2)}, \ldots, \vec{f}_{i}^{(M)}\right\}$.

\subsubsection{Attention layer}
The attention mechanism focuses the model's attention on key information \cite{luong2015effective}. We use the attention mechanism to fuse the user features with the tweet features. The inputs of this layer are the user features $H$ from the GAT layer and the text features $F$ from the BERT. The users and the texts are not in one-to-one correspondence, but one user corresponds to multiple texts. Since the final task is at the text level, only the user features corresponding to the text are retained in the current batch. Then the user features are appended onto the text features, denoted as $X_{i}=\left\{\vec{f}_{i}^{(1)}, \ldots, \vec{f}_{i}^{(M)}, \vec{h}_{i}^{\prime(1)}, \ldots, \vec{h}_{i}^{\prime(N)}, \vec{r}_{i}\right\}$. In addition, to reinforce the location information of the sequence, we inject the position encoding according to the method in Transformer \cite{vaswani2017attention}. Note that the user features from different attention heads of GCNs use the same position encoding. The resulting features are multiplied by the three weight matrices $W$ to obtain the matrices $Q$, $K$, $V$ after performing layer normalization \cite{ba2016layer}, and the attention is computed as follows.

\begin{equation}
	Attention(Q, K, V)=softmax\left(\frac{Q K^{T}}{\sqrt{d_{k}}}\right) V
\end{equation}

\begin{equation}
	head_{i}=Attention\left(Q W_{i}^{Q}, K W_{i}^{K}, V W_{i}^{V}\right)
\end{equation}

\begin{equation}
MultiHead(Q, K, V)=Concat\left(head_{1}, \ldots, head_{h}\right) W^{O}
\end{equation}

Where $d_{k}$ is the dimensionality of matrix $K$ and $W_{i}^{Q}$, $W_{i}^{K}$, $W_{i}^{V}$, $W_{i}^{O}$ is the parameter matrix. The output of multi-headed attention is denoted as $X_{i}^{\prime}=\left\{\vec{x}_{1}, \ldots, \vec{x}_{M}, \vec{x}_{M+1}, \ldots \vec{x}_{M+N+1}\right\}$. After this, we add residual linking and layer normalization.

\subsubsection{Feed-Forward layer}
Following the attention layer is a feedforward network layer, which contains a ReLU activation function and a linear transformation.
\begin{equation}
	FFN(x)=\max (0, x) W+b
\end{equation}

Where $W$ is the weight of the fully connected layer and $b$ is the bias.

\subsubsection{Loss function}
For the data imbalance problem, we use Focal Loss \cite{lin2017focal} to mitigate the classification difficulty of imbalanced samples.
\begin{equation}
	L_{f l}=\left\{\begin{array}{cc}
		-\alpha\left(1-y^{\prime}\right)^{\gamma} \log y^{\prime}, \quad y=1 \\
		-(1-\alpha) y^{\prime \gamma} \log \left(1-y^{\prime}\right), y=0
	\end{array}\right.
\end{equation}

Where  $\alpha$ and $\gamma$ are both adjustable hyperparameters, we set $\alpha=0.25$ and $\gamma=2$. The $y^{'}$ is the model prediction value, which is between $(0-1)$. When $y=1$, $y^{'}$ tends to $1$, indicating easy positive samples, and its contribution to the weights tends to $0$. When $y=0$, $y^{'}$ tends to $0$, indicating easy negative samples, and its contribution to the weights tends to $0$.

\section{Experiments and Results}
In this section, we evaluate the performance of the proposed method. First, we describe the experimental setup. Second, our method is compared with the current state-of-the-art methods. Third, we evaluate the effectiveness of the modules in CT-OLD. Then we compare the performance of feature initialization methods. Finally, we explore the differences introduced by different social graph modeling methods.

\subsection{Experimental setup}
To evaluate the effectiveness of the end-to-end detection method proposed in this paper for offensive language detection, we conducted a series of experiments. Our method uses a batch size of 64, and the rest of the methods are executed according to the original papers. All experiments are conducted on 12,780 samples, 1,260 users, and 8,877 relations. The dataset is divided 7:3 by tweets and the average results are calculated by 10 independent replications. All experiments are executed on a server with 128G RAM, Intel Xeon Gold 6130 2.10GHz, and NVIDIA Tesla V100. In experiments, all models are trained with maximum 20 epochs. Besides, the early stop strategy is used if the F1 score doesn't increase in five continuous epochs.

\subsubsection{Baseline analysis}
To perform performance evaluation, we conducted a series of baseline studies of these state-of-the-art methods for offensive language detection as follows.

\textbf{BERT.} Liu et al. \cite{liu2019nuli} won the first place in subtask A of OffensEval-2019 with the fine-tuned BERT model. The text embeddings are obtained by the BERT model and then classified by linear layer.

\textbf{BERT+CNN.} Safaya et al. \cite{safaya2020kuisail} combined BERT with CNN for offensive language detection. Their method ranked in the top four on all three languages in subtask A of OffensEval-2020. BERT+CNN learns text representations through the BERT model. Then the output of the last four layers of BERT is fed into CNN networks with different size filters.

\textbf{LR+AUTH.} \footnote{The code of this baseline is taken from \url{https://github.com/pushkarmishra/AuthorProfilingAbuseDetection}.} LR+AUTH \cite{mishra2018author} is the first work to combine community structure and texts for offensive speech detection. First, social relationships are constructed as social graphs. Then user representations are learned by node2vec. For each tweet, the user features are appended to the n-gram features of the tweet, and the LR classifier is trained.

\textbf{LR+GCN.} LR+GCN \cite{mishra2019abusive} models users and tweets to construct a heterogeneous graph. It uses bags of words as the feature of users. This method uses two layers of GCNs to train by minimizing the cross-entropy of the labeled nodes in the graph. The output of the first hidden layer of the trained GCNs is used as the user embeddings. For each tweet, user features are appended onto the character n-gram features of the tweet for training the LR classifier. State-of-the-art results were obtained at that time.

The LR+AUTH code is publicly available and directly used in our experiments. The rest of the baseline is our implementation based on the original paper.

\subsubsection{Performance metrics}
To quantify model effects, we show AUC, Accuracy, Macro Recall, Macro Precision, and Macro F1 score.

\begin{equation}
	A U C=\frac{\sum_{i \in  positiveclass} rank_{i}-\frac{M \cdot(M+1)}{2}}{M \cdot N}
\end{equation}

Where $\sum_{i \in  positiveclass} rank_{i}$ is the sum of the positive samples ranked by probability score from smallest to largest, $M$ is the number of positive samples, and $N$ is the number of negative samples. When the probability scores are the same, the $rank$ value is the average of the $rank$ of the same probability scores.

For the other four metrics, the confusion matrix is calculated first, and for each classification, the true positive (TP), the false negative (FN), the false positive (FP), and the true negative (TN) are calculated as follows:

\begin{equation}
	Accuracy=\frac{T P+T N}{T P+T N+F P+F N}
\end{equation}

\begin{equation}
	Recall_{i}=\frac{T P_{i}}{T P_{i}+F N_{i}}
\end{equation}

\begin{equation}
	Precision_{i}=\frac{T P_{i}}{T P_{i}+F P_{i}}
\end{equation}

\begin{equation}
	F 1_{i}=\frac{2 \cdot  Precision_{i} \cdot Recall_{i}}{Precision_{i}+Recall_{i}}
\end{equation}

The subscript $i$ of each metric represents the $i$-th category, and the number of categories is represented by $L$. After calculating the Recall, Precision, and F1 for each category, the average value is used to obtain the macro value. The formula for calculating Macro-F1 is given below, and the rest of the metrics are the same.

\begin{equation}
	Macro -F 1=\frac{1}{L} \sum_{i=1}^{L} F 1_{i}
\end{equation}

\subsection{Evaluation of the proposed detection approach}
\subsubsection{Comparison with baseline method}

\renewcommand{\arraystretch}{1.5}
\begin{table}[htbp]
	\centering
	\fontsize{7}{8}\selectfont
	\caption{The baselines vs. CT-OLD}
	\setlength{\tabcolsep}{0.5mm}{
	\begin{tabular}{@{}cccccc@{}}
		\toprule
		\textbf{Model} &  \textbf{Accuracy} & \textbf{Precision} & \textbf{Recall} & \textbf{F1} \\ \midrule
		LR+AUTH\cite{mishra2018author}      &  93.87 & 82.97 & 71.57 & 75.80    \\
		LR+GCN\cite{mishra2019abusive}  &  95.62 & 89.70 & 79.23 & 83.49    \\
		BERT\cite{liu2019nuli}  &  95.41 & 83.70 & 88.80 & 86.02    \\
		BERT+CNN\cite{safaya2020kuisail}    & 95.64 & 84.65 & 88.65 & 86.51    \\
		CT-OLD(Ours)     &  \textbf{96.92}    & \textbf{89.83}     & \textbf{90.06}  & \textbf{89.94}    \\ \bottomrule
	\end{tabular}}
	\label{baselines}
\end{table}

\begin{figure}
	\centering
	\includegraphics[scale=0.55]{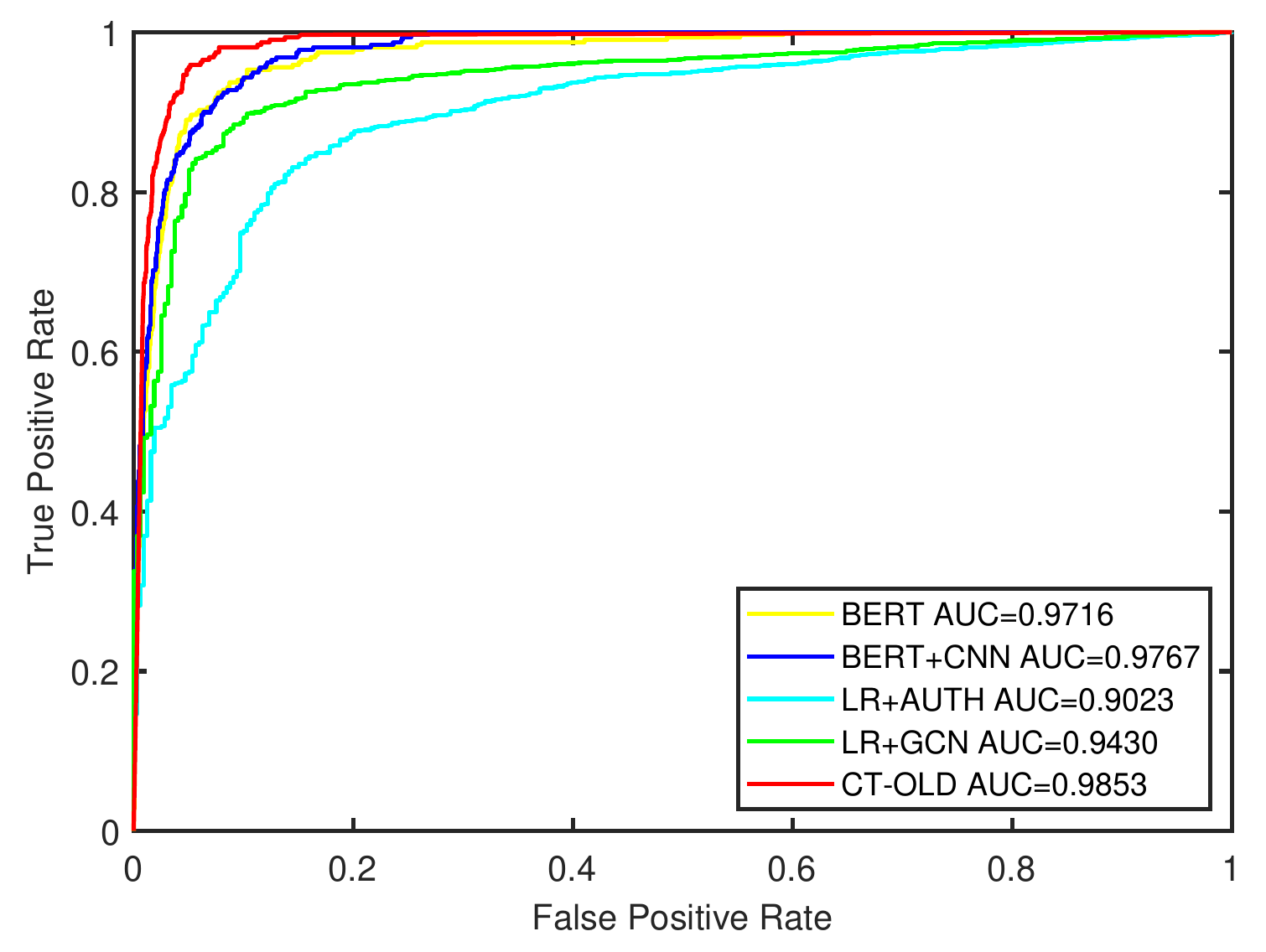}
	\caption{ROC-AUC of baselines and CT-OLD}
	\label{auc}
\end{figure}

Table \ref{baselines} shows the AUC, Accuracy, Precision, Recall, and F1 score of the baseline methods and the CT-OLD on our dataset. In LR+AUTH and LR+GCN, text features are extracted by n-grams, resulting in relatively poor performance. In addition, they are classified after calculating user embeddings and text embeddings separately. This divides the training process into two parts. LR+AUTH learns user embeddings only through community structure and has the worst performance. LR+GCN considers user features represented by text, which greatly improves the performance of detection. The BERT-based method obtains competitive scores by considering only text. And BERT-CNN has further improvement, but it is not significant.

CT-OLD obtains the first place in all evaluation metrics, and the F1 score was 3.43\% higher than the second. For the ROC curve, we find that CT-OLD's curve covers all other methods. This proves that our method has a great advantage in the offensive language detection task.

\subsubsection{Effect of different training ratios}

This experiment shows the performance of the baseline method with CT-OLD at different training ratios from 0.1 to 0.9.

\begin{figure*}[htbp]
	\centering
	\includegraphics[scale=0.35]{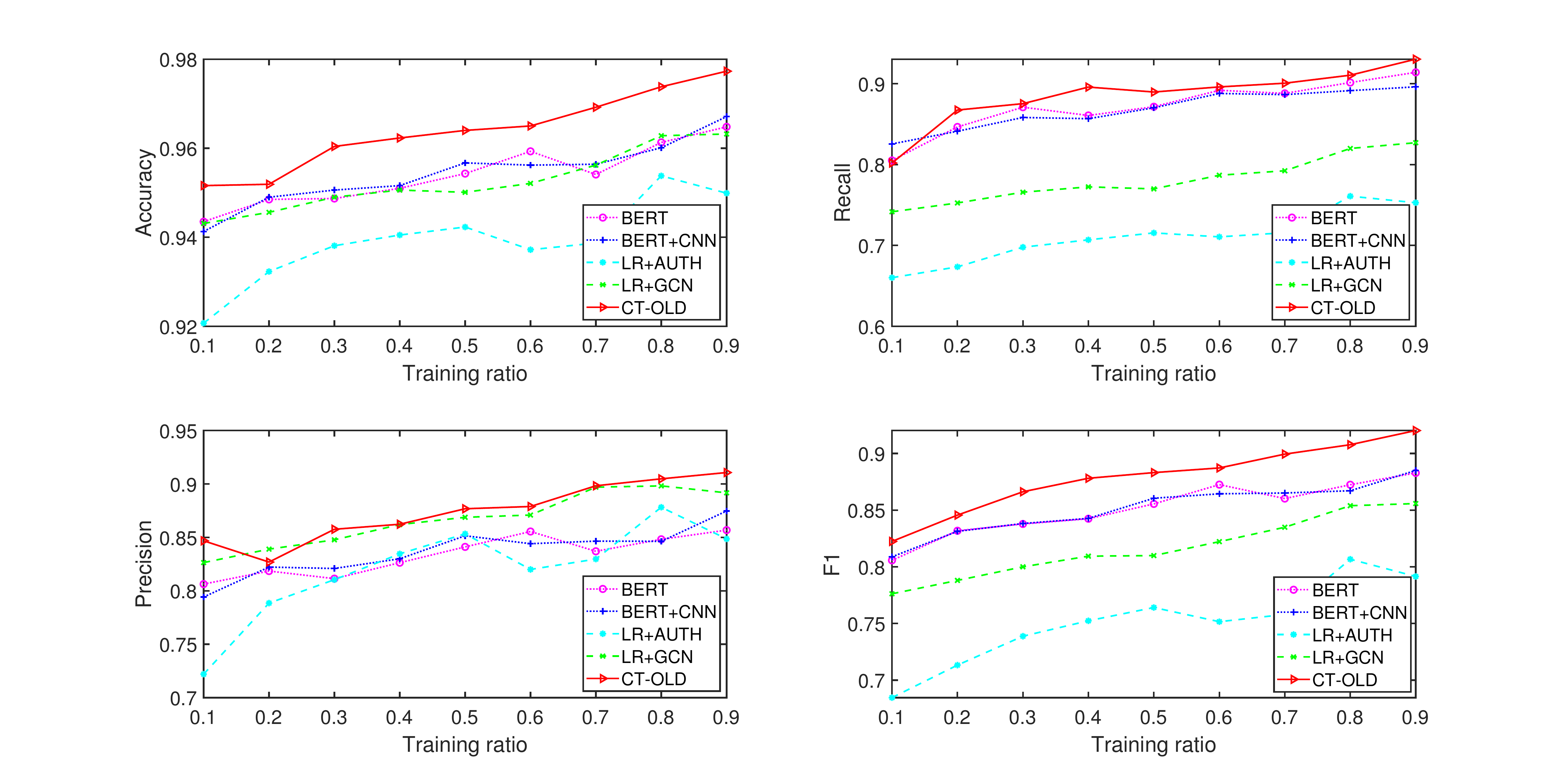}
	\caption{Accuracy, Precision, Recall, and F1 of baselines and CT-OLD for training sets with percentages between 0.1 and 0.9}
	\label{train}
\end{figure*}

It can be seen from Figure \ref{train} that different training ratios have an impact on all methods. Meanwhile, each performance metric gradually increases with the increase of the training ratio. The difference between BERT and BERT-CNN in different training ratios is small. The F1 score of LR+AUTH fluctuates the most. And it is lower than the other methods at all ratios.

The F1 score of CT-OLD increases more obviously when the training ratio is from 0.1 to 0.4. In addition, the difference between the F1 scores of CT-OLD and BERT is also increasing. This is because when the training set is small, the node features are relatively sparse on the social graph. With increasing the training data, the opinion of users to post offensive language is learned obviously. As a whole, CT-OLD achieves a significant advantage in all metrics.

\subsubsection{Detection model ablation tests}

To explore the influence of the graph attention network layer, the BERT, the residuals of the GAT layer, the multi-headed attention of the GAT layer, and the attention layer used in the CT-OLD, we perform ablation tests. Specifically, we remove only one of the above modules from the CT-OLD and show the performance of the model. The results are shown in Table \ref{ablation}

\renewcommand{\arraystretch}{1.5}
\begin{table}[htbp]
	\centering
	\fontsize{7}{8}\selectfont
	\caption{The impact of modules on detection performance}
	\setlength{\tabcolsep}{0.1mm}{
	\begin{tabular}{@{}cccccc@{}}
		\toprule
		\textbf{Model} &
		\textbf{AUC} &
		\textbf{Accuracy} &
		\textbf{Precision} &
		\textbf{Recall} &
		\textbf{F1} \\ \midrule
		CT-OLD\textbackslash{}GAT             & 0.9775 & 95.83 & 87.02 & 84.90 & 85.92 \\
		CT-OLD\textbackslash{}BERT            & 0.9225 & 92.44 & 75.48 & 85.33 & 79.31 \\
		CT-OLD\textbackslash{}Residual of GAT & 0.9802 & 96.35 & 88.18 & 87.75 & 87.96 \\
		\begin{tabular}[c]{@{}c@{}}CT-OLD\textbackslash{}Multi-head \\ Attention of GAT\end{tabular} & 0.9824 &	96.50 &	87.63 &	90.40 &	88.95 \\
		CT-OLD\textbackslash{}Attention Layer & 0.9852 & 96.56 & 88.13 & 89.86 & 88.97 \\
		CT-OLD &	\textbf{0.9853} &	\textbf{96.92} &	\textbf{89.83} &	\textbf{90.06} &	\textbf{89.94} \\ \bottomrule
	\end{tabular}}
	\label{ablation}
\end{table}

We find that the removal of any module will reduce the model performance. All modules contribute to the effectiveness of the CT-OLD. The CT-OLD$\backslash$GAT removes the graph attention network layer, with similar results to the two BERT-based methods used as baselines. The CT-OLD$\backslash$BERT removes the text embeddings obtained by BERT. The connection between the GAT and our detection target is only the user historical behavior information. Meanwhile, it only detects the granularity of the author. In such a coarse-grained detection, the F1 score has reached 79.31\%, exceeding the LR+AUTH. The GAT uses community structure with user behavior information rather than textual content. So, the information it learns is different from the text-based method. In addition, the text-based BERT captures semantic information well. Therefore, the end-to-end detection method combining the two features significantly improves the detection performance.

In the remaining four experiments, the CT-OLD$\backslash$Residual of GAT removes the residual links from the graph attention network and has the most degraded performance. This is because the residual connection enhances the influence of the original features.

\subsubsection{Comparison of initialization of user features}

The node features are an important factor for graph representation learning. However, for users whose tweets exclude in the training set, their initial features are not available during the training. This requires feature initialization for this subset of users. Therefore, we consider four initialization methods.
\begin{enumerate}[(1)]
	\item All-0 initialization. The number of tweets for both categories is unknown. It is set to $[0,0]$ by default.
	\item All-1 initialization. The probability of a user posting tweets in both categories is the same. Set to $[1,1]$.
	\item Average initialization. For each category of tweets, the average number of user posts is used as the initial value.
	\item Non-offensive initialization. There are a few users who post offensive language in social networks. And it is assumed that users only post non-offensive tweets. It is set to $[1,1e-6]$.
\end{enumerate}
In addition to the default case of 0.7 training ratio, we also show the detection performance in the sparse data case of 0.1 training ratio. Table \ref{init} shows the performance of different feature initialization methods.

\renewcommand{\arraystretch}{1.5}
\begin{table}[htbp]
	\centering
	\fontsize{7}{8}\selectfont
	\caption{Performance of different feature initialization methods}
	\setlength{\tabcolsep}{0.5mm}{
	\begin{tabular}{@{}cccccc@{}}
		\toprule
		\textbf{training ratio} & \textbf{Model} & \textbf{Accuracy} & \textbf{Precision} & \textbf{Recall} & \textbf{F1} \\ \midrule
		\multirow{4}{*}{Sparse data} & All-0   & 94.17          & 80.51          & 77.85          & 79.11          \\
		& All-1    & 94.60          & 81.76          & \textbf{80.44} & 81.08          \\
		& Avg    & 94.70          & 82.79          & 78.79          & 80.63          \\
		& Non-off & \textbf{95.16} & \textbf{84.68} & 80.19          & \textbf{82.24} \\ \midrule
		\multirow{4}{*}{Default} & All-0   & 96.45          & 87.05          & 91.22          & 89.00          \\
		& All-1    & 96.40          & 86.79          & \textbf{91.34} & 88.90          \\
		& Avg    & 96.61          & 87.70          & 91.31          & 89.40          \\
		& Non-off & \textbf{96.92} & \textbf{89.83} & 90.06          & \textbf{89.94} \\ \midrule
	\end{tabular}}
	\label{init}
\end{table}

The effectiveness of user historical behavior information decreases in the case of sparse data. The detection performance is more dependent on the user feature initialization. The experiments show that non-offensive initialization performs the best. By default, only a small part of users rely on feature initialization. The performance of the four methods is similar. The difference between the highest and lowest F1 score is only less than 1\%.

\subsubsection{Comparison of social graph modeling}

In the past, modeling of community information could be done by computing node embeddings through the network structure of the communities \cite{mishra2018author}. This method is able to capture the structure of communities but ignores the association with offensive language detection. 

Mishra et al. \cite{mishra2019abusive} built user features through the n-gram features of tweets, which effectively combined community structure and text information. In this method, community structure and linguistic features are captured together. The user features obtained by this method contain all the information of the tweets posted by the user. However, the user representations of this method depend on the quality of the tweet representations deeply.

We compare three social graph modeling methods for users.
\begin{enumerate}[(1)]
	\item Using the modeling method of Mishra et al. \cite{mishra2019abusive}, but we only model the users. The user features are constructed by binary bag-of-words representations of the user's posted tweets. This method is called the bow feature.
	\item We tag users based on whether they post offensive language. If the user has posted offensive language, the tag is 1, otherwise, it is 0. This method we call the hard feature.
	\item In our proposed method, we use the user historical behavior information as user features. This method is called the soft feature.
\end{enumerate}

\renewcommand{\arraystretch}{1.5}
\begin{table}[htbp]
	\centering
	\fontsize{7}{8}\selectfont
	\caption{The performance of different social graph modeling methods}
	\setlength{\tabcolsep}{0.5mm}{
	\begin{tabular}{@{}cccccc@{}}
		\toprule
		\textbf{Model}     & \textbf{Accuracy} & \textbf{Precision} & \textbf{Recall} & \textbf{F1} \\ \midrule
		GAT-bow     & 92.33 & 75.20 & 84.56 & 78.88 \\
		GAT-hard    & 85.08 & 66.75 & 87.02 & 70.57 \\
		GAT-soft    & 92.44 & 75.48 & 85.33 & 79.31 \\ \midrule
		CT-OLD-bow   & 96.01 & 86.15 & 88.70 & 87.37 \\
		CT-OLD-hard  & 96.35 & 88.42 & 87.32 & 87.86 \\
		CT-OLD-soft      & \textbf{96.92}    & \textbf{89.83}     & \textbf{90.06}  & \textbf{89.94}     \\ \bottomrule
	\end{tabular}}
	\label{modeling}
\end{table}

Table \ref{modeling} shows the comparison in different methods of social graph modeling. Among the three methods of GAT, the F1 score using the soft feature is the highest, the bow feature is the second-highest, and the hard feature is the lowest. This may be because the hard feature contains too little information. As long as users have posted offensive language, their tweets are judged to be offensive. The bow feature contains complete information, but large-scale features also have more uncertainty during training. The soft feature has more constraints than the bow feature and more complete information than the hard feature.

The soft feature still gets the highest F1 score among the three CT-OLD methods, while the difference between the bow feature and hard feature is not obvious. The reason for this variation is that the graph learning module using the hard feature and soft feature has a clear purpose to learn the representation of users to post offensive language. The bow feature, on the other hand, uses raw features of text embeddings with complex features, resulting in difficulties in learning effective information. Therefore, the hard feature has stronger performance than the bow feature. However, the hard feature will be mislabeled if the training set excludes the offensive tweets of a user and the test set does, affecting the detection performance. The soft feature based on user historical behavior information has the advantages of others and achieves better performance.

\section{Conclusion}

In this paper, we proposed an end-to-end method for detecting offensive language with graph attention networks and BERT. The model captures community structure and text information at the same time and fuses them by attention mechanisms.  Specifically, we used user historical behavior information to represent user opinion. This information is a node attribute on the social graph that associates the social graph with the offensiveness of the user. We evaluated the effectiveness of our method on the dataset we constructed. The experimental results showed that the proposed method is effective for detecting offensive language, and it has better performance than other methods.

In online social networks, the generalization ability of the offensive language detection model that learns community structure and text at the same time will exceed that of the text-based model. But it is difficult to obtain more datasets with different user distribution and community structure. In the future, we hope to further investigate the impact of social graphs on generalization ability.

\bibliography{mybibfile}

\end{document}